\input{aipcheck}

\documentclass[
    ,final            
  ]
  {aipproc}

\layoutstyle{6x9}

\usepackage{xspace}

\def\cref#1{Chapt.\,\ref{#1}}
\def\Cref#1{Chapter~\ref{#1}}

\def\fref#1{Fig.\,\ref{#1}}
\def\ffref#1{Figs.\,\ref{#1}}

\def\rref#1{Ref.\,\cite{#1}}

\def\lleft{\textit{left}\xspace}
\def\rright{\textit{right}\xspace}
\def\LLeft{\textit{Left}\xspace}
\def\RRight{\textit{Right}\xspace}

\def\line{---}
\def\dashed{-\,-\,-}
\def\dotted{$\cdot\cdot\cdot$}
\def\dashdot{-$\cdot$-$\cdot$}

\def\gcm2{g/cm$^2$\xspace}
\def\Cerenkov{\v{C}erenkov\xspace}

\def\Section#1{\section{#1}}

\def\hh{}
\def\fwidth{0.99\textwidth}

\begin{document}

\title{The Composition of Cosmic Rays at the Knee
\footnote{Invited talk, given at the Centenary Symposium 2012: Discovery of
Cosmic Rays, June 2012, Denver.}}


\author{J\"org R.\ H\"orandel}{
  address={Department of Astrophysics/IMAPP, Radboud University Nijmegen,\\
    P.O. Box 9010, 6500 GL Nijmegen, The Netherlands ~---~
    http://particle.astro.ru.nl}
}

\begin{abstract}
 Recent results on the composition of cosmic rays in the energy region from
 about $10^{14}$ to $10^{18}$~eV are reviewed.
\end{abstract}

\maketitle


\Section{Introduction}
The all-particle energy spectrum of cosmic rays (ionized atomic nuclei from
outer space) follows a broken power law $dN/dE\propto E^\gamma$. At an energy
around $4\cdot 10^{15}$~eV the spectral index changes from $\gamma\approx-2.7$
at low energies to $\gamma\approx-3.1$.  This structure in the energy spectrum
is referred to as "the knee". Understanding the physics behind this "kink" in
the energy spectrum is thought to be crucial to understand the origin of
Galactic cosmic rays \cite{behreview}.

The origin of cosmic rays is threefold: particles with energies below about
100~MeV are accelerated during solar bursts in magnetic reconnections at the
surface of the Sun. Nuclei with energies up to several $10^{17}$~eV are usually
considered to originate within our Milky Way, most likely being accelerated in
Supernova Remnants. At the highest energies ($>10^{18}$eV), particles are
usually considered to be of extragalactic origin.
In the following, we will focus on the composition and origin of Galactic
cosmic rays, i.e.\ cosmic rays with energies between $\approx1$ and
$10^{9}$~GeV.

At energies around 1~GeV individual elements have been identified in direct
measurements of cosmic rays above the atmosphere, all elements from the
periodic table of elements are present in cosmic rays, e.g.\ \cite{cospar06}.
Their abundance follows roughly the abundance in the local Galactic
environment/solar system.  Energy spectra for the major cosmic-ray elements
from hydrogen to iron have been directly measured up to energies approaching
$10^{14}$~eV (e.g.\ Fig. 26.1 in \rref{Beringer:1900zz}). They follow power
laws with a spectral index close to $\gamma\approx-2.7$. Below about 10~GeV the
spectra are influenced by solar modulation, i.e.\ deflection/attenuation in the
Heliosphere.

The cosmic-ray flux is strongly falling as a function of energy.  Thus, at
energies exceeding $10^{14}$~eV large areas are needed to collect a
statistically significant data sample in a reasonable time. The large areas
needed, exceeding the size of a soccer field, can only be realized on ground,
i.e.\ below the atmosphere of the Earth. Thus, secondary particles are measured
which are produced in extensive air showers, initiated by high-energy cosmic
rays impinging onto the atmosphere.

The intrinsic fluctuations during the development of a particle cascade limit
the mass resolution of experiments registering air showers. Typical
mass-sensitive observables in such experiments are the depth of the shower
maximum in the atmosphere $X_{max}$ and the ratio of electrons to muons at
ground level $N_e/N_\mu$.  Using a Heitler type model for hadronic showers
\cite{matthewsheitler} it can be shown that these observables depend on the
nuclear mass number of the primary particle $A$ as $X_{max}\propto -X_0 \ln A$
and $\lg(N_e/N_\mu)\propto -0.065\ln A$, with the radiation length in air $X_0$
\cite{ricap07,jrherice06}.  Thus, if one requires e.g.\ a resolution of
$\Delta\ln A\approx1$, the experimental accuracies have to be better than
$\Delta X_{max}\approx 36$~\gcm2 or $\Delta(N_e/N_\mu)\approx 16\%$. These
values are close to the experimental uncertainties of current air shower
experiments. In state-of-the-art air-shower experiments a resolution of the
order of $\Delta\ln A\approx0.8$ has been achieved, i.e.\ the major cosmic-ray
elements from hydrogen to iron can be divided in up to 5 elemental groups
(instead of 26 individual elements as registered in direct measurements above
the atmosphere).

\Section{Recent Results}
\begin{figure}[t]
 \includegraphics[width=0.5\textwidth]{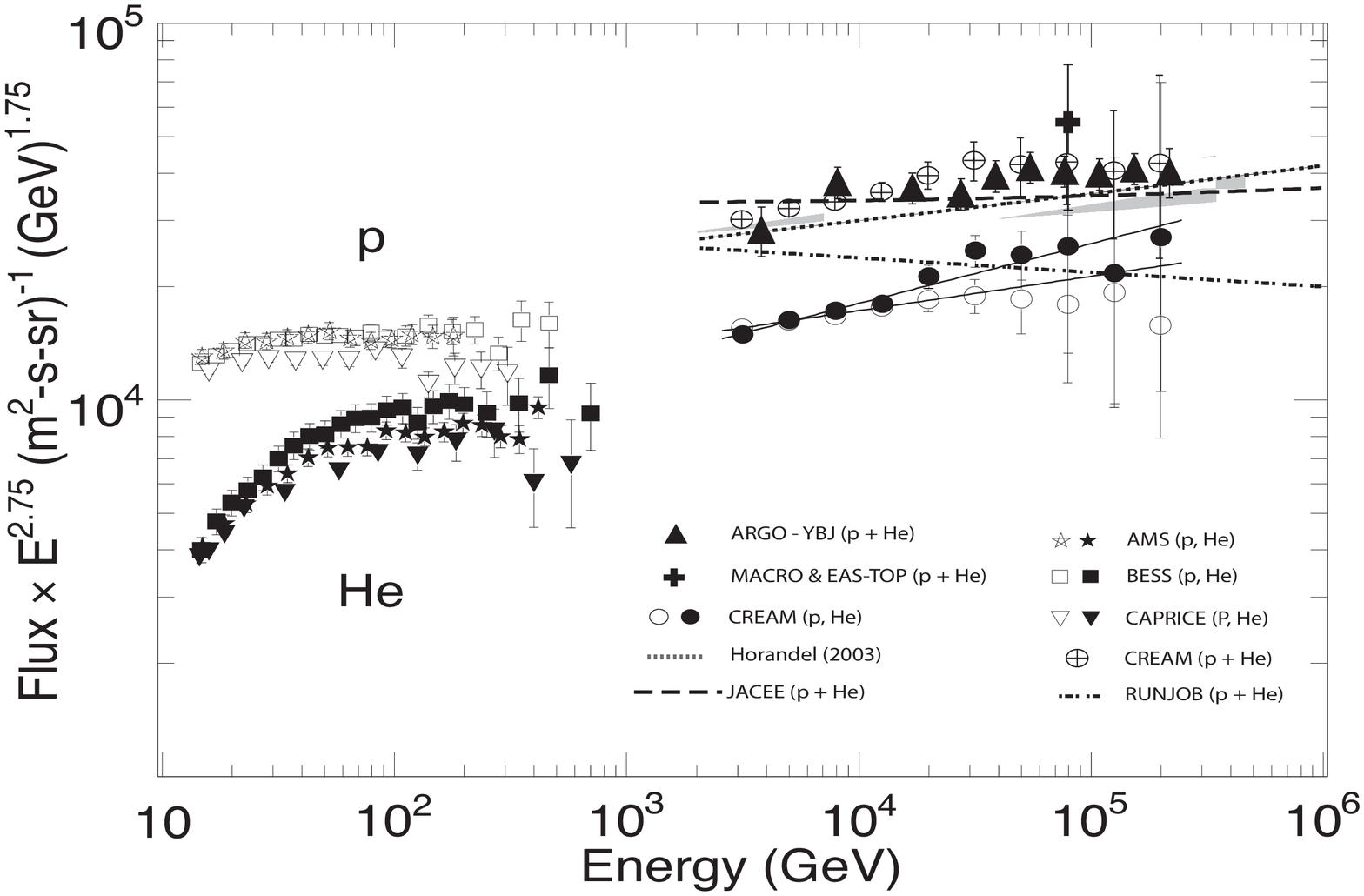}
 \includegraphics[width=0.5\textwidth]{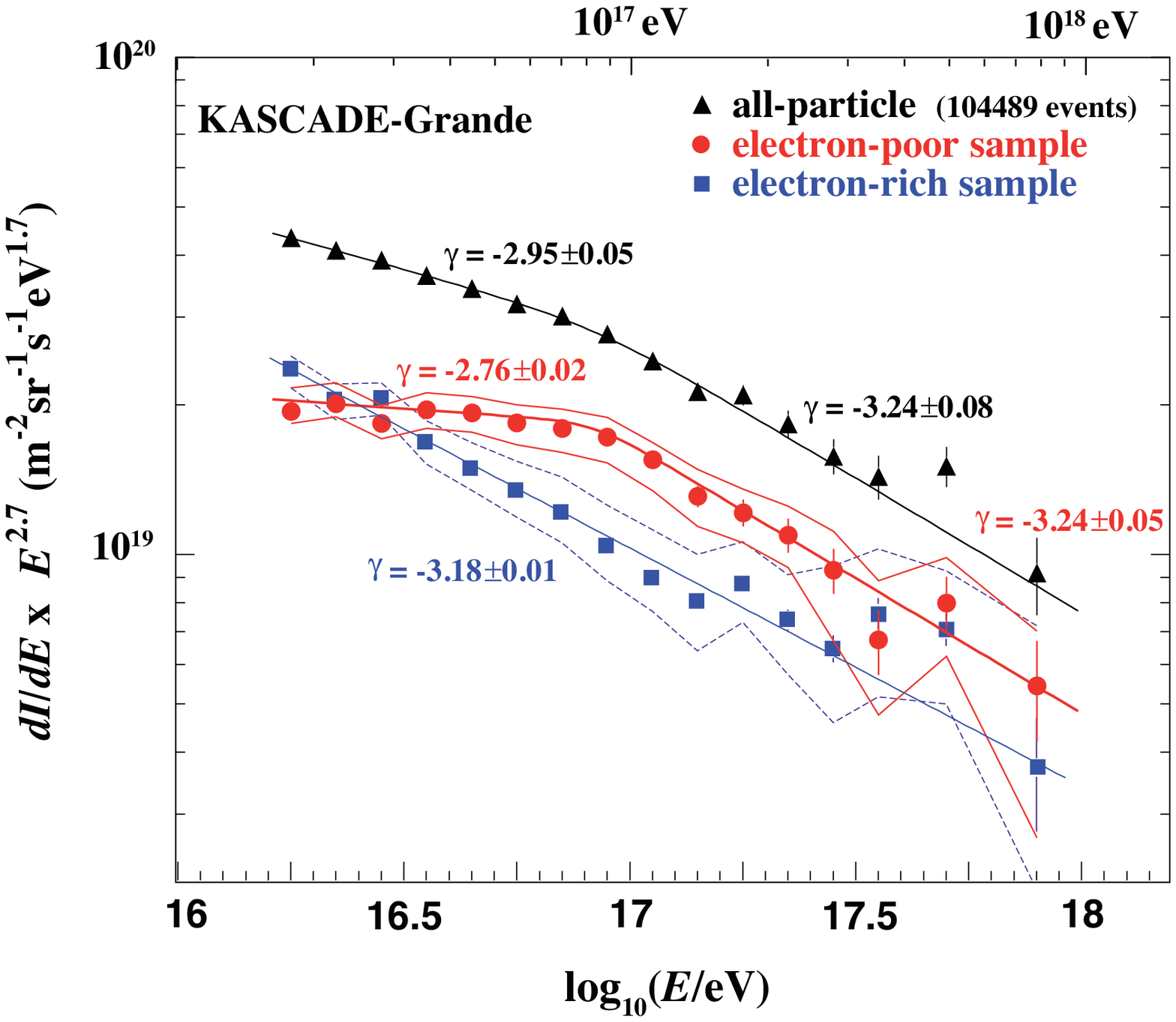}
  \caption{\LLeft: Energy spectrum of the "light" cosmic-ray component (p+He),
             as measured by ARGO-YBJ \cite{argoicrc32}.
	   \RRight: Energy spectrum of cosmic rays as determined with
	     KASCADE-Grande: all-particle spectrum (black)
             and heavy component (red) \cite{Apel:2011mi}.\hh}
  \label{argokg}
\end{figure}

\paragraph{ARGO-YBJ}
The ARGO-YBJ experiment is situated at the Yangbajing International Cosmic-Ray 
Observatory, located at an altitude of about 4300 m a.s.l.\ in Tibet
\cite{argoicrc32}.  The detector consists of an air-shower array, made of a
full-coverage RPC carpet.  The energy spectrum of the light cosmic-ray
component (protons and helium) has been measured in the energy range from 1 to
300~TeV, as depicted in \fref{argokg} (\lleft).  Results of the CREAM balloon
experiment are in good agreement with the air-shower data from ARGO. The latter
are also compatible with a parameterization of the world data \cite{pg}, as can
be inferred from the figure. This agreement in the energy region of overlap
between direct and indirect cosmic-ray measurements is a valuable test of the
simulation codes used to interpret the air shower data.

\paragraph{KASCADE(-Grande)}
The KASCADE experiment, located in Germany measured simultaneously all three
air shower components \cite{kascadenim}.  The number of electrons and muons in
air showers has been used to unfold the energy spectra for five groups of
elements \cite{ulrichapp,Apel:2008cd}. The results indicate that the knee is
caused by the fall-off of the energy spectrum of protons, followed by
subsequent cutoffs of heavier elements, approximately proportional to the
rigidity ($E/Z$) of the individual elements.
This observation has been confirmed by independent measurements, using a muon
tracking detector to determine the average production depth of muons in the
atmosphere \cite{Apel:2011zz}.

The experiment has been extended to cover an area of about 0.5~km$^2$ ---
KASCADE-Grande \cite{Apel:2010zz}. The all-particle energy spectrum has been
measured in the energy range from $10^{16}$ to $10^{18}$~eV, derived from
measurements of the total number of charged particles and the total muon number
of muons \cite{Apel:2012rm}. The energy spectrum exhibits strong hints for a
hardening of the spectrum at approximately $2\cdot 10^{16}$~eV and a
significant steepening at $8\cdot10^{16}$~eV.
The data have also been used to divide the primary cosmic rays in a light and a
heavy group and to determine the respective energy spectra, as shown in
\fref{argokg} (\rright) \cite{Apel:2011mi}.  One recognizes a fall-off of the
heavy component at an energy around $8\cdot10^{16}$~eV.  Such a behavior is
expected, assuming a rigidity dependent cut-off of the individual cosmic-ray
elements.

Based one these measurements, one expects that the average composition becomes
heavier as a function of energy up to about $10^{17}$~eV and again lighter
above.  Such a behavior is confirmed by the following two observations.

\begin{figure}[t]
 \includegraphics[width=0.5\textwidth]{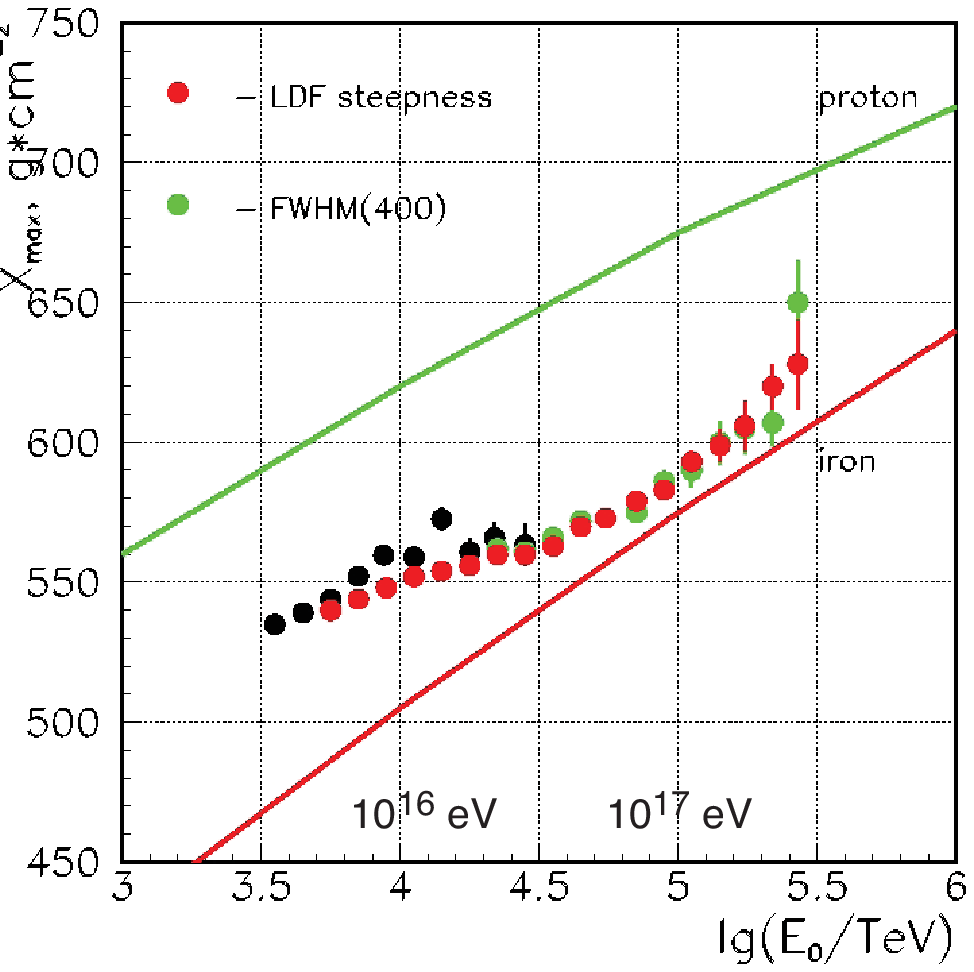}
 \includegraphics[width=0.5\textwidth]{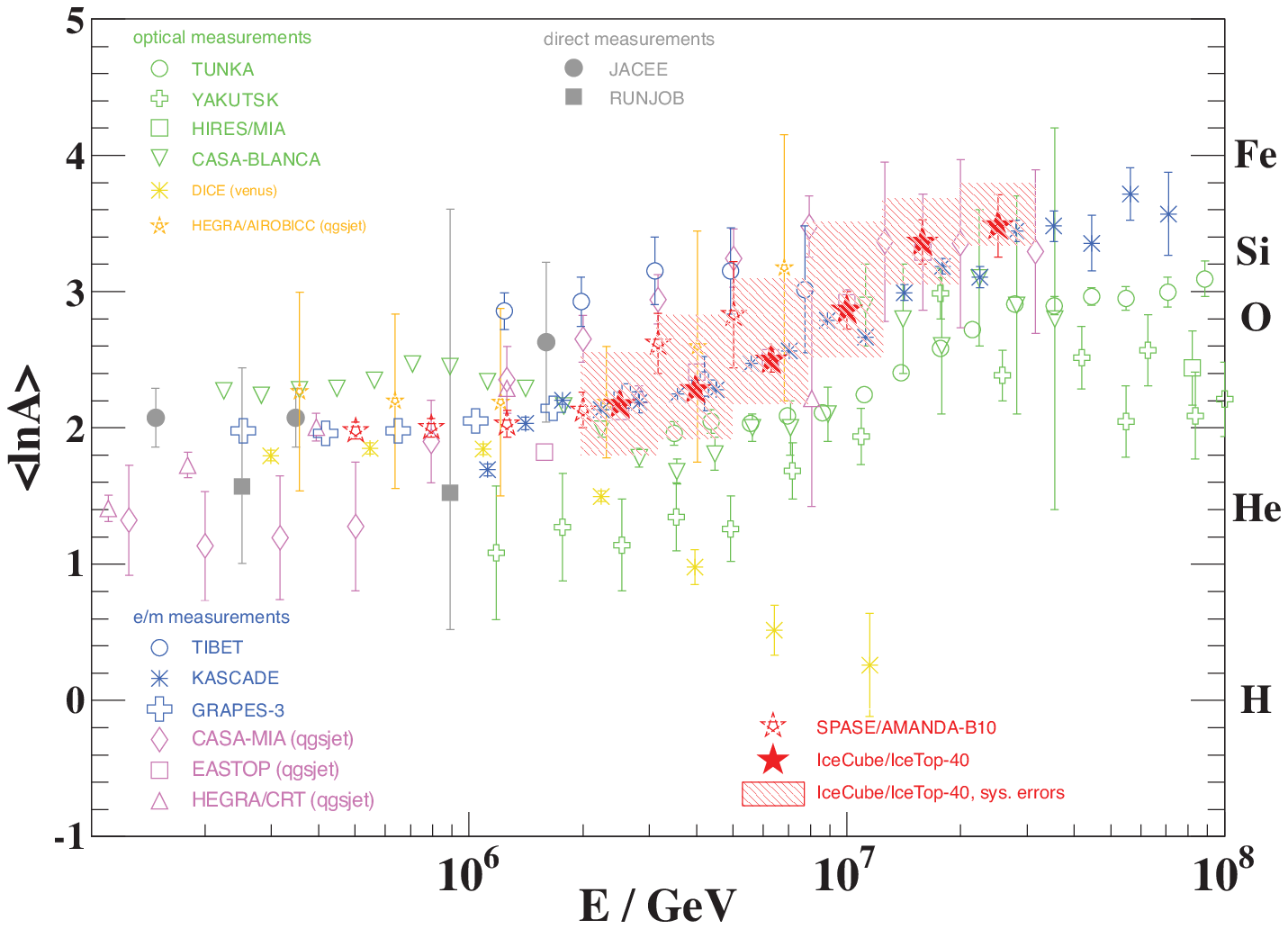}
  \caption{\LLeft: Average depth of the shower maximum as a function of energy,
             measured by Tunka \cite{tunkaicrc32}.
           \RRight: Mean logarithmic mass as a function of energy obtained by
             IceCube/IceTop \cite{Tamburro:2012ay}.\hh}
  \label{tunkaice}
\end{figure}

\paragraph{Tunka}
The Tunka-133 air shower array with about 1~km$^2$ geometric acceptance area is
installed in the Tunka Valley (50 km from Lake Baikal) \cite{tunkaicrc32}. It
measures the \Cerenkov component of air showers with an array of 133 open
\Cerenkov detectors.  The average depth of the shower maximum as measured by
Tunka is presented in \fref{tunkaice} (\lleft). The lines represent
predictions for primary protons and iron nuclei. It can be recognized that as
a function of energy, the average cosmic-ray composition becomes heavier
(approaching iron) up to about $10^{17}$~eV and again lighter at higher
energies.

\paragraph{IceCube/IceTop}
The IceCube neutrino telescope at the South Pole is also a large km$^2$-scale
detector for muons from extensive air showers, complemented by an array of
detectors on the surface (IceTop) to register the charged particles in air
showers \cite{Tamburro:2012ay}.
First results on the mass composition of cosmic rays have been obtained from
data taken already during the construction of the detector.  The mean
logarithmic mass derived from one month of data with about half the detector is
depicted in \fref{tunkaice} (\rright).
The measurements clearly indicate a rising mean mass as a function of energy.
Results up to energies exceeding $10^{17}$~eV are expected soon with the full
detector being operational since 2010 and it will be interesting to see, if a
trend to a lighter composition, as discussed above, will be found as well by
IceCube at energies exceeding $10^{17}$~eV.

\Section{The Composition of Galactic Cosmic Rays}
\begin{figure}[t]
 \includegraphics[width=\fwidth]{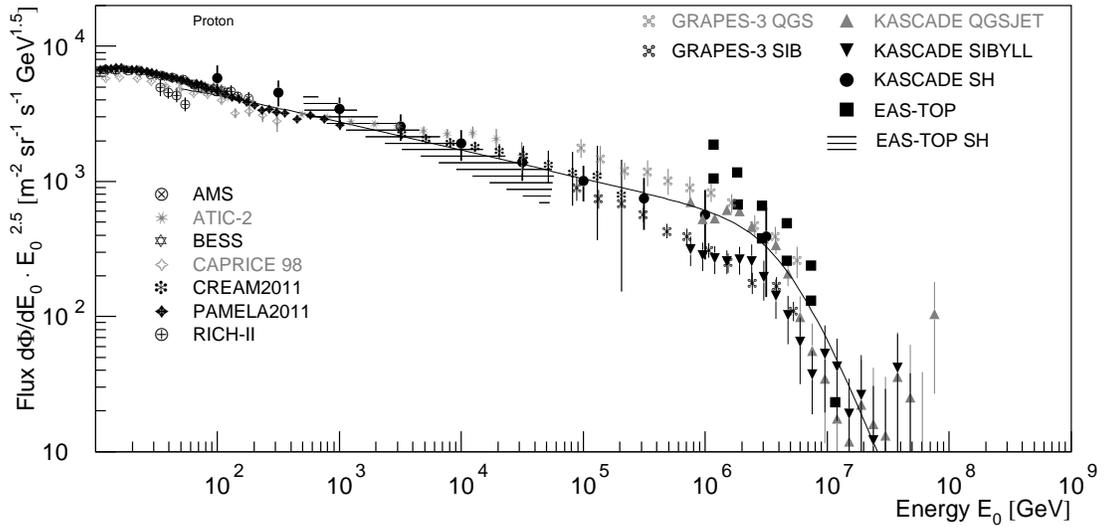}
  \caption{Energy spectrum of protons in Galactic cosmic rays as measured by
     the direct experiments
     AMS \cite{amsp},
     ATIC-2 \cite{Panov:2011ak},
     BESS \cite{bess00},
     CAPRICE~98 \cite{caprice98},
     CREAM2011 \cite{Yoon:2011zz},
     PAMELA2011 \cite{Adriani:2011cu}, and
     RICH-II \cite{rich2}
     and the air shower experiments
     GRAPES \cite{grapes05},
     EAS-TOP (electrons and muons) \cite{Ghia:2005gn}
             (unaccompanied hadrons) \cite{eastopsh}, and
     KASCADE (electrons and muons) \cite{ulrichapp} 
             (unaccompanied hadrons) \cite{kascadesh}.
     Two hadronic interaction models have been used to interpret the data
     from the GRAPES and KASCADE experiments.
     The line represents a parameterization according to the Poly Gonato model
     \cite{pg}.\hh}
  \label{protons}
\end{figure}

\begin{figure}[t]
 \includegraphics[width=\fwidth]{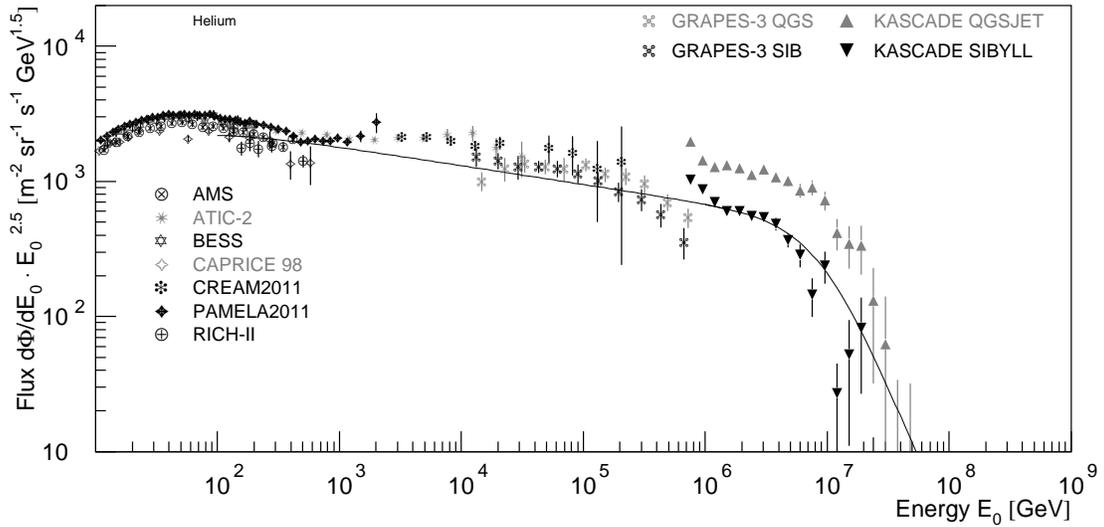}
  \caption{Energy spectrum of helium nuclei in Galactic cosmic rays as measured
     by the direct experiments
     AMS \cite{amsp},
     ATIC-2 \cite{Panov:2011ak},
     BESS \cite{bess00},
     CAPRICE~98 \cite{caprice98},
     CREAM2011 \cite{Yoon:2011zz},
     PAMELA2011 \cite{Adriani:2011cu}, and
     RICH-II \cite{rich2}
     and the air shower experiments
     GRAPES \cite{grapes05}, and
     KASCADE \cite{ulrichapp}.
     Two hadronic interaction models have been used to interpret the data
     from the GRAPES and KASCADE experiments.
     The line represents a parameterization according to the Poly Gonato model
     \cite{pg}.\hh}
  \label{helium}
\end{figure}

\begin{figure}[t]
 \includegraphics[width=\fwidth]{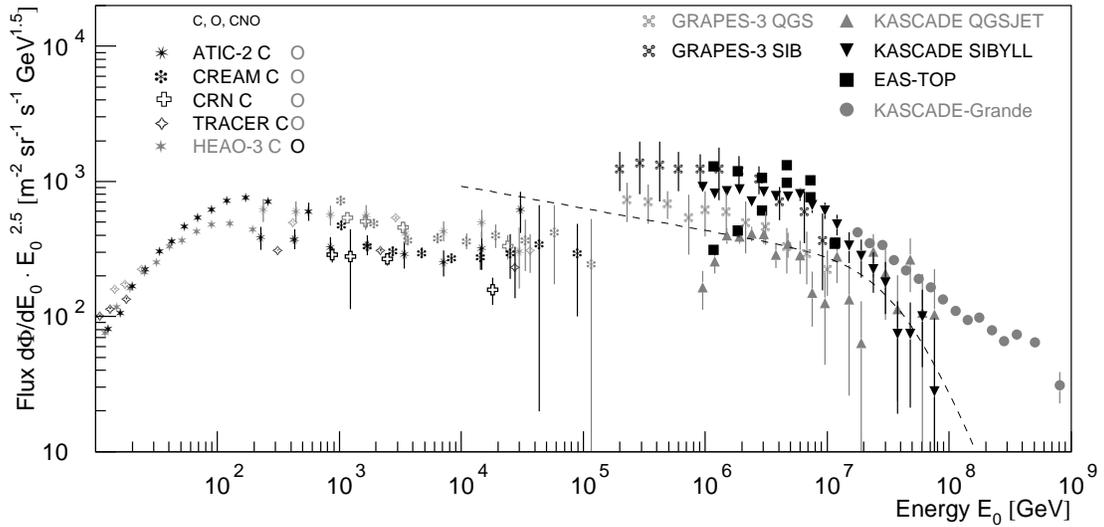}
  \caption{Energy spectrum of nuclei from the CNO group
     in Galactic cosmic rays as measured by the direct experiments
     ATIC-2 \cite{Panov:2011ak},
     CREAM \cite{Yoon:2011zz},
     CRN \cite{crn},
     HEAO-3 \cite{heao3},
     TRACER \cite{tracer03}
     and the air shower experiments
     GRAPES \cite{grapes05},
     EAS-TOP \cite{Ghia:2005gn},
     KASCADE \cite{ulrichapp}, and 
     KASCADE-Grande (light) \cite{Apel:2011mi}.       
     Two hadronic interaction models have been used to interpret the data
     from the GRAPES and KASCADE experiments.
     The direct measurements have single-element resolution, i.e.\ measure the
     flux of carbon and oxygen nuclei. Air shower experiments can only resolve
     elemental groups.
     The line represents a parameterization according to the Poly Gonato model
     \cite{pg}.\hh}
  \label{CNO}
\end{figure}

\begin{figure}[t]
 \includegraphics[width=\fwidth]{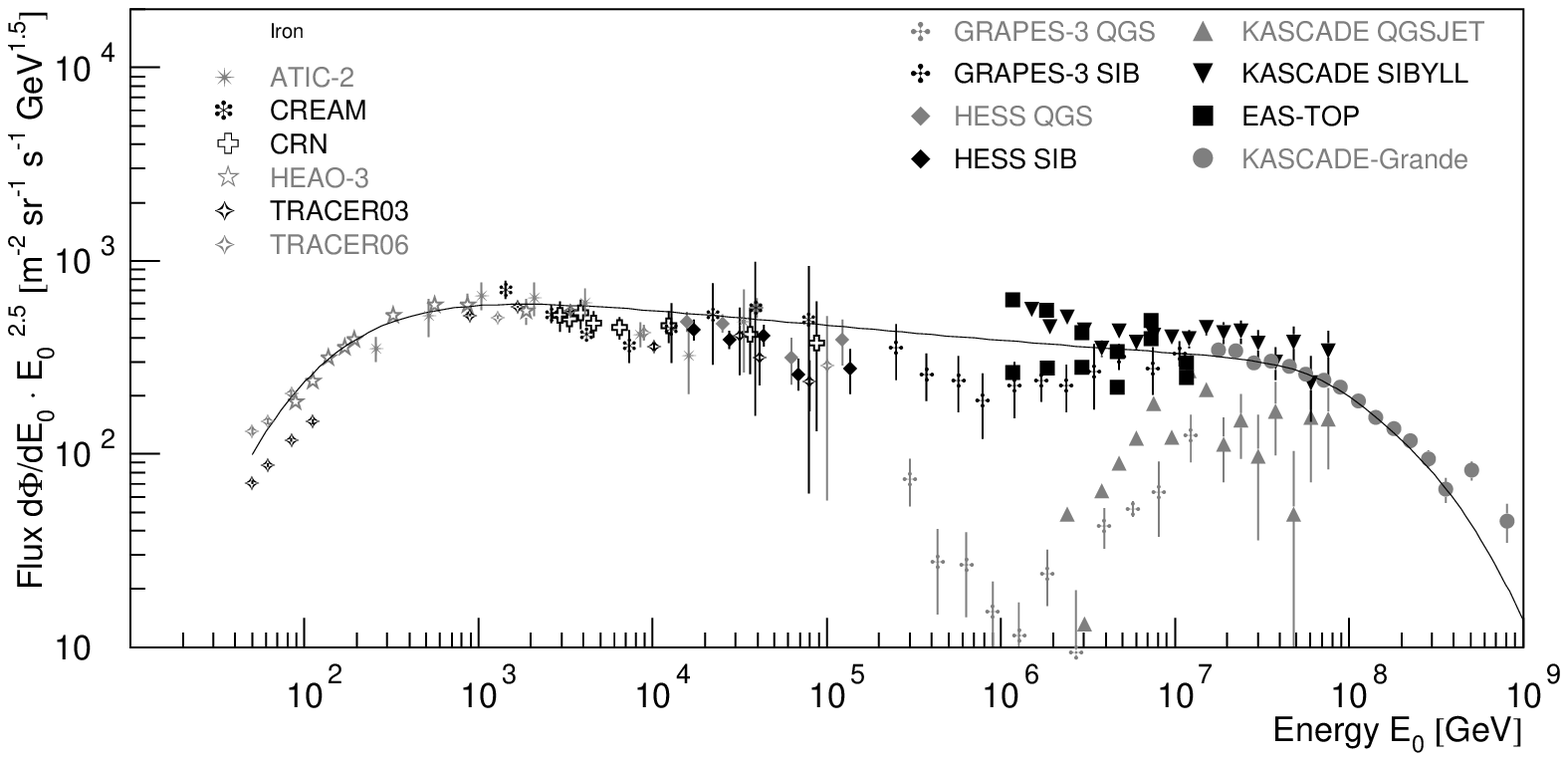}
  \caption{Energy spectrum of nuclei from the iron group
     in Galactic cosmic rays as measured by the direct experiments
     ATIC-2 \cite{Panov:2011ak},
     CREAM \cite{Yoon:2011zz},
     CRN \cite{crn},
     HEAO-3 \cite{heao3},
     TRACER 2003 \cite{tracer03} and 2006 \cite{Obermeier:2011wm}
     and the air shower experiments
     GRAPES \cite{grapes05},
     H.E.S.S. direct \Cerenkov light \cite{Aharonian:2007zja},
     EAS-TOP \cite{Ghia:2005gn},
     KASCADE \cite{ulrichapp}, and 
     KASCADE-Grande (heavy) \cite{Apel:2011mi}.       
     Two hadronic interaction models have been used to interpret the data
     from the GRAPES and KASCADE experiments.
     The direct measurements have single-element resolution, i.e.\ measure the
     flux of iron nuclei. Air shower experiments can only resolve
     elemental groups.
     The line represents a parameterization according to the Poly Gonato model
     \cite{pg}.\hh}
  \label{iron}
\end{figure}

A compilation of world data from direct and indirect measurements of cosmic
rays for four elemental groups is given in \fref{protons} (protons),
\fref{helium} (helium nuclei), \fref{CNO} (CNO-group nuclei), and
\fref{iron} (iron-group nuclei).  Here we restricted ourself to "modern"
measurements. Older data are included in previous compilations
\cite{pg,cospar06}.
The energy is given as total energy per particle.
Direct measurements above the atmosphere (on balloons and space crafts) extend
to almost $10^6$~GeV and at higher energies air shower measurements set in.

To guide the eye the lines represent a parameterization according to the
Poly Gonato model with a rigidity dependent cut-off and a constant
$\Delta\gamma$ (see \rref{pg} for details) with the following parameter range
for the nuclear charge number $Z$:
\fref{protons} protons $Z=1$,
\fref{helium} helium $Z=2$,
\fref{CNO} CNO group $Z=5-12$,
\fref{iron} iron group $Z=26-92$.

These figures reflect the present status of our understanding of the elemental
composition of Galactic cosmic rays. Several common features can be recognized.
At low energies, the flux is influenced by magnetic fields in the heliosphere
(solar modulation).
At higher energies the spectra follow approximately a power law.
Finally, at energies exceeding $10^{15}$~eV the spectra exhibit a fall-off,
which is roughly proportional to the charge of the respective nuclei
$E_c\approx Z \cdot 4\cdot10^{15}$~eV.

A closer look reveals some more properties.
An often discussed issue is the spectral slope of protons and helium nuclei. As
can be inferred from \ffref{protons} and \ref{helium}, the spectrum of helium
is slightly flatter ($\gamma=-2.64\pm0.02$) as compared to protons
($\gamma=-2.71\pm0.02$).
Looking closely at the proton and helium spectra, a structure might be visible
around 200~GeV. Above this energy, the spectra follow power laws, which extend
into the air-shower energy region, where ultimately a cut-off is observed.
Below about 200~GeV, both proton and helium exhibit a "bump", before the solar
modulation yields to a depression at lowest energies.
This feature is sometimes referred to as "spectral hardening"
\cite{Yoon:2011zz}. However, from \ffref{protons} and \ref{helium} it appears
as there are two cosmic-ray components, one below 200~GeV and a second one at
higher energies.
It should also be noted that the effect is very subtle and one may ask if
systematic effects in the experiments are understood to such a precision, in
particular, since the energy corresponds roughly to the transition between two
experimental techniques: from magnet spectrometers (at low energies) to
calorimeters.

Looking at the CNO and iron groups, it may be noted that the recent
KASCADE-Grande data (\fref{argokg}, \rright) extend previous measurements to
higher energies and a cut-off is now also clearly visible in the iron group.
Since protons and helium nuclei have already reached their cut-off energies,
the "light" component in \fref{argokg}, \rright corresponds most likely to the
CNO group.
An interesting behavior can be observed for the iron group: two hadronic
interaction models (QGSJET and SIBYLL) have been used to interpret the air
shower data measured by GRAPES and KASCADE. For the interaction model QGSJET a
"dip" is visible for both experiments at energies around $10^6$~GeV.  To derive
the spectra the correlations between the number of electrons and muons in the
showers are investigated.  QGSJET is not compatible with the measured
distributions around energies of $10^6$ GeV. This yields the depression in the
iron spectrum, consistently observed by both experiments.  Such a behavior has
been observed earlier, for a detailed discussion see also \rref{ulrichapp} and
\cite{Apel:2008cd}.
It might also be worth to mention that the recent KASCADE-Grande data for the
heavy/iron component are right on top of the predictions of the Poly Gonato
model (published a decade before the measurements).

\Section{Hadronic Cross Sections and LHC Data}

Air showers measured with the Pierre Auger Observatory have been used to derive
the proton-air interaction cross section at a center-of-mass energy of 57~TeV
\cite{Collaboration:2012wt}.
For the proton-air
cross section at ($57\pm6$)~TeV a value of
$$ \sigma_{p-air}=\left[505\pm22 (\mbox{stat}) ^{+20}_{-15} (\mbox{syst})
\right]~\mbox{mb} $$
has been derived.  Using Glauber theory a value for the inelastic proton-proton
cross section has been derived:
$$\sigma_{pp}=\left[92\pm7 (\mbox{stat}) ^{+9}_{-11} (\mbox{syst}) \pm 7
(\mbox{Glauber}) \right]~\mbox{mb}. $$

\begin{figure}[t]
 \includegraphics[width=0.5\textwidth]{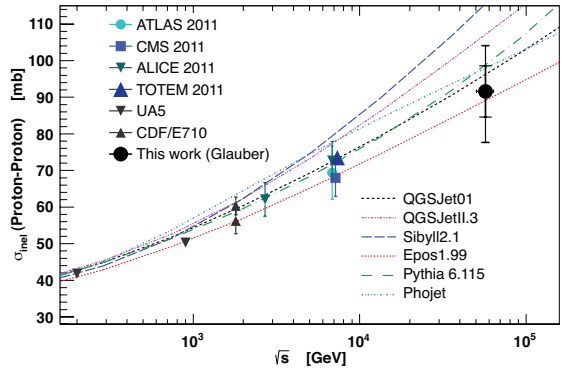}
  \caption{Inelastic proton-proton cross section as a function of the
   center-of-mass energy as measured by various accelerator experiments and
   recently derived from air shower data by the Pierre Auger collaboration
   \cite{Collaboration:2012wt}. The lines represent predictions from different
   hadronic interaction models.\hh}
  \label{fig1}
\end{figure}

The Auger value is shown together with accelerator measurements, including
recent LHC data  in \fref{fig1}. Also shown are predictions from different
hadronic interaction models.  It should be noted, that the recent Auger results
and the recent LHC measurements are at the lower boundary of the range
predicted by the various interaction models.

The recent results indicate lower values for the proton-air and proton-proton
cross sections.  A lower cross section has already been predicted in 2003
\cite{wq}. In this work the mean logarithmic mass derived from air shower
observations has been systematically investigated. In particular, a discrepancy
has been found between experiments observing the depth of the shower maximum
and experiments registering the number of secondary particles at ground.  It
has been shown that a smaller proton-proton and consequently also proton-air
cross section reduces the discrepancy in the derived mean logarithmic mass from
the different classes of observables.  The recent results confirm this earlier
findings and this example illustrates how the new LHC measurements directly
influence the interpretation of air shower data.

In the past, various ideas have been discussed as possible origin of the knee
in the energy spectrum of cosmic rays, among them were ideas about new types of
(hadronic) interactions in the atmosphere, see e.g.\ \cite{origin}.  The recent
LHC results exhibit a fair agreement with predictions of hadronic interaction
models using standard physics, see e.g.\ \cite{pierogbeijing,jrhrap}. Hence, it
can be concluded that the knee in the energy spectrum is not caused by new
physics in the atmosphere, it is rather of astrophysical origin.

\Section{The Origin of Galactic Cosmic Rays}

\begin{figure}[t]
 \includegraphics[width=\fwidth]{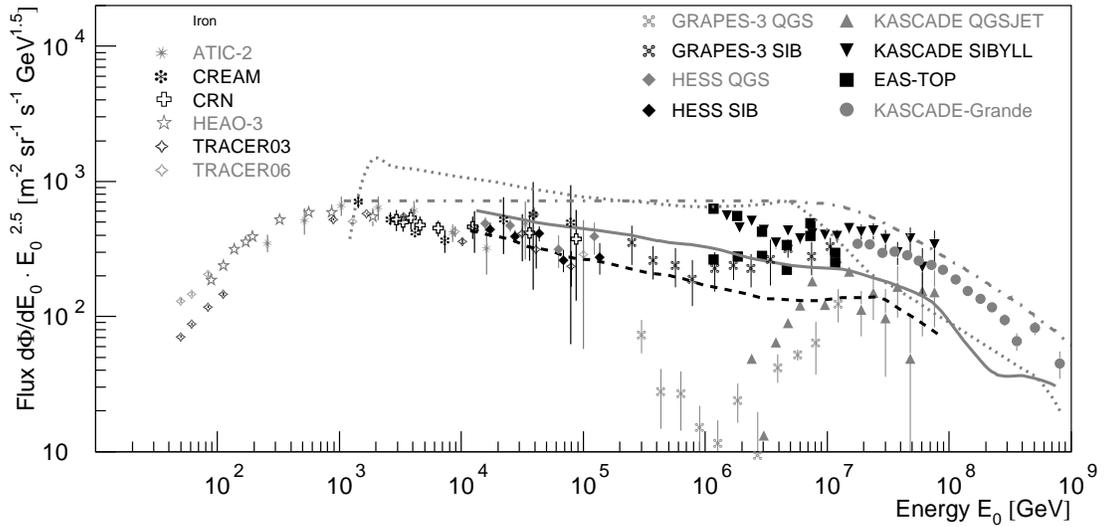}
  \caption{Energy spectrum of the cosmic-ray iron group.
     Experimental data as in \fref{iron}.
     The lines indicate spectra for models explaining the knee due to the
     maximum energy attained during the acceleration process according to
   Sveshnikova et al. \cite{sveshnikova} (\line),
   Berezhko et al. \cite{berezhko} (\dashed),
   Stanev et al. \cite{stanev} (\dotted), and
   Kobayakawa et al. \cite{kobayakawa} (\dashdot).\hh}
  \label{snr}
\end{figure}

The origin of the knee in the energy spectrum is often assumed to give hints
towards the origin of Galactic cosmic rays.  In the literature various
scenarios are discussed as possible origin for the knee. They can be grouped in
four classes \cite{origin}: (i) the finite energy reached during the
acceleration process (presumably in Supernova Remnants), (ii) leakage from the
Galaxy during the diffusive propagation process, (iii) interaction with
background particles during the propagation process, and (iv) new hadronic
interactions within the atmosphere, which transport a fraction of the energy
into unobserved channels.

The third class can most likely be excluded due to to measurements of the
composition of cosmic rays in the energy region around $10^{16}$ to
$10^{17}$~eV \cite{origin}. The interactions with background particles would
result in a  break-up of heavy nuclei, in which many light particles, such as
protons and $\alpha$ particles are produced. Thus, on expects a rather light
elemental composition in the energy range mentioned. However, recent data
indicate a heavy composition in this energy regime, see e.g.\ \fref{tunkaice}.

Also the fourth class can be disregarded due to recent measurements.
Air shower data from the KASCADE experiment have been used to
check the predictions of hadronic interaction models. A quantitative analysis
demonstrates, that the predictions of the models agree with measured
observables on the 10\% level \cite{jenspune}.
A big step forward in the understanding of hadronic interactions 
are the results from the Large Hadron Collider (LHC). The design center-of-mass energy of 14~TeV
corresponds to a laboratory energy of $10^{17}$~eV, i.e.\ well above the knee
in the energy spectrum.
The recent LHC results exhibit a fair agreement with predictions of hadronic
interaction models using standard physics, see e.g.\
\cite{pierogbeijing,jrhrap}. 
All this information together leaves very little room for new type of
interactions inside the atmosphere.  Thus, it can be concluded that the knee in
the energy spectrum is not caused by new physics in the atmosphere, it is
rather of astrophysical origin.

\begin{figure}[t]
 \includegraphics[width=\fwidth]{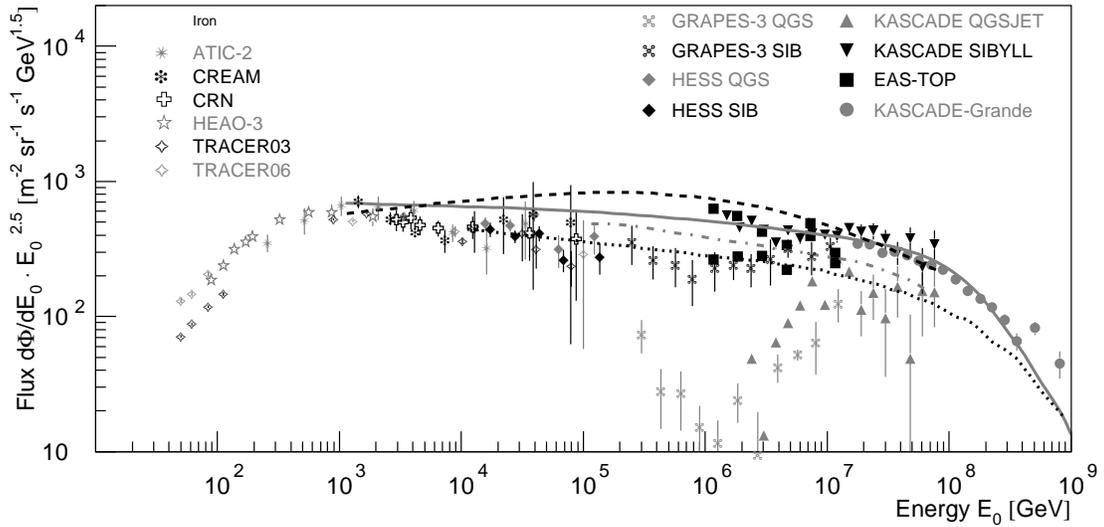}
  \caption{Energy spectrum of the cosmic-ray iron group.
     Experimental data as in \fref{iron}.
   The lines indicate spectra for models explaining the knee as an 
   effect of the propagation process according to
   H\"orandel \& Kalmykov et al. \cite{prop} (\line),
   Ogio et al. \cite{ogio} (\dashed),
   Roulet et al. \cite{roulet} (\dotted), as well as
   V\"olk et al. \cite{voelk} (\dashdot).\hh}
  \label{prop}
\end{figure}

Predictions of astrophysical models are illustrated in \ffref{snr} and
\ref{prop} for the cosmic-ray iron group.  A comparison of the model
predictions with measured data for the light groups is given in
\rref{cospar06}.
Models describing the acceleration of cosmic rays in Supernova Remnants predict
a maximum energy attained during the acceleration process, which is
proportional to the nuclear charge number of the respective atomic nuclei $Z$
and the strength of the magnetic field $B$ in the acceleration region
$E_{max}\propto Z\cdot B$. The predictions of such models are depicted in
\fref{snr}.
A second class of models describes the diffusive propagation in the Galaxy. In
this models one finds a maximum rigidity $E/Z$ above which the nuclei cannot be
any more effectively magnetically bound to the Galaxy and the particles leak
out of the Milky Way. The predictions of such models are shown in \fref{prop}.
For a more detailed discussion of the models the reader is referred to
\rref{origin}. 

The different models exhibit some scatter, however, the magnitude of the
differences is of the same order as the accuracy of the measurements. A general
trend can be observed: all models shown describe well the fall-off of the iron
group as seen in the measurements.
I would like to put some emphasize on the predictions of \rref{prop}.
It has been shown that the propagation process alone results in a very soft
bend of the individual spectra, not compatible with the measured strong fall-off
(see \ffref{protons} to \ref{iron}). In addition, a rigidity dependent maximum
energy in the sources had to be included to describe the measured sharpness of
the cut-offs.
Thus, I would like to conclude that the origin of the knee in the energy
spectrum is a combination of the maximum energy attained during the
acceleration process and leakage from the Galaxy.
This fits very well with the "standard picture" of Galactic cosmic rays, being
accelerated in Supernova Remnants and diffusively propagating through the Milky
Way.


\begin{theacknowledgments}
 The author thanks his colleagues from TRACER, KASCADE-Grande, and the
 Pierre Auger Observatory for fruitful discussions.
\end{theacknowledgments}

\bibliographystyle{aipproc}   


\end{document}